\documentclass[aps,reprint,prl,groupedaddress,showkeys]{revtex4-1}

\usepackage{graphicx}

\begin{document}


\title{New insights into the photochromic mechanism in oxygen-containing yttrium hydride thin films: an optical perspective} 


\author{J. Montero}
\thanks{jose.montero.amenedo@ife.no}
\author{F. A. Martinsen}
\author{S. Zh. Karazhanov}
\author{B. C. Hauback}
\author{E. S. Marstein}
\affiliation{Institute for Energy Technology, P.O. Box 40, NO-2027 Kjeller (Norway)}


\date{\today}

\begin{abstract}

Oxygen-containing yttrium hydride thin films exhibit a  photochromic behavior: transparent  thin films switch from a transparent state to a photodarkened state after being illuminated with UV or blue light. This feature has attracted much attention in  recent years due to its potential applications  in  smart fenestration or in any device in which a response to intense light radiation is needed. However, the process responsible for the reversible change of the optical properties upon illumination is still not well understood. The objective of the present work is to shed some light on the photochromic mechanism by using an optical approach. On this basis, the optical properties of oxygen-containing yttrium hydride thin films have been studied by optical spectrophotometry and ellipsometry before (transparent state) and after UV illumination (dark state). According to the observed results, the photochromic optical change of the films can be explained quantitatively by the gradual growth, under illumination, of metallic phases within the initial wide-band gap semiconducting lattice. 

\end{abstract}.  

\maketitle 


Photochromic yttrium hydride thin films have promising technological applications in smart windows, optical switches, optical storage, sensors or in any other device in which a response to  intense light radiation is required. In a window, these coatings can play an important role, both regulating the energy throughput and reducing the glare, enhancing the thermal and luminous comfort of the users \cite{YoshimuraLanghammerDam2013,GranqvistGreenNiklassonEtAl2010,BaetensJelleGustavsen2010,Lampert2004}.

The reversible change of the optical properties of yttrium after hydrogenation was first reported by Huiberts \emph{et al.} \cite{Huiberts1996}; in short this mechanism consists of the treatment of a metallic yttrium layer  with hydrogen in order to form an  yttrium hydride. As a metallic Y thin film is hydrogenated, a face centered cubic (\emph{fcc}) YH$_2$ phase is formed initially. Both  Y and YH$_2$  have a metallic-like optical behavior and therefore are opaque to visible light. However, further hydrogenation of YH$_2$ results in the formation of hexagonal close packed (\textit{hcp}) YH$_3$, which is a wide band gap semiconducting phase, and therefore transparent in the visible spectrum. The phase transformation from \emph{fcc} to \emph{hcp} in YH$_x$ films is known to take place around x = 2.8 \cite{OhmuraMachidaWatanukiEtAl2007}. Besides, when the YH$_3$ films are subjected to high pressure ($\sim$ 20 GPa), a transition from the \textit{hcp} YH$_3$ phase to the \emph{fcc} YH$_3$ phase takes place, causing the gap closure of YH$_3$ and hence a sudden drop of the optical transparency \cite{WijngaardenHuibertsNagengastEtAl2000, MachidaOhmuraWatanukiEtAl2006, PalasyukTkacz2005,AlmeidaKimOrtizEtAl2009}.

In addition, yttrium hydride films also exhibit  photochromic (PC) behavior, i.e., the optical properties of the films change reversibly when illuminated by light of adequate energy (wavelengths in the blue or UV range). Early works by Hoekstra \emph{et al.} \cite{HoekstraRoyRosenbaumEtAl2001} have shown a light induced metal insulator transition in yttrium hydrides at low temperature, and Ohmura \emph{et al.} \cite{OhmuraMachidaWatanukiEtAl2007,OhmuraMachidaWatanukiEtAl2006} accidentally discovered PC behavior in yttrium hydride  films subjected to high pressure. Later, Mongstad \emph{et al.} \cite{MongstadPlatzer-BjoerkmanMaehlenEtAl2011,MongstadPlatzer-BjoerkmanKarazhanovEtAl2011} reported PC  behavior in transparent oxygen-rich yttrium hydride  films under atmospheric conditions and at room temperature. In the latter case, however, the yttrium hydride films were directly obtained by reactive magnetron sputtering rather than by the subsequent hydrogenation of a pre-deposited metallic Y layer. 

The mechanism of the PC behavior in yttrium hydride is still unclear and seems to have no relation with the PC mechanism reported for transition oxides \cite{ChandranSchreudersDamEtAl2014}. 
In the present work, the PC behavior is studied using systematic optical measurements supported by theoretical modeling.
In particular,  the optical properties  of oxygen-rich transparent semiconducting thin films -- hereafter referred in the text simply as YH$_x$ -- in both their clear and darkened state, have been investigated by ellipsometry and spectrophotometry. In addition, the optical properties of opaque metallic thin films -- from now referred in the text simply as YH$_y$, where $y<x$ -- have been also studied. Both sets of films, YH$_x$ and YH$_y$, were obtained by sputter deposition at a hydrogen/argon ratio $\Gamma=$ 0.18 and 0.13 respectively and then exposed to air where they oxidize.  
The optical approach to the study of the PC mechanism is deliberate; the biggest change observed in the films under illumination takes place in the optical transmittance, while only small changes are observed in the microstructural analysis \cite{MaehlenMongstadYouEtAl2013}. The modeling of the experimental results revealed that the optical properties of the YH$_x$ (photodarkened) films can be quantitatively explained by the formation of  metallic YH$_y$ domains embedded into the YH$_x$  (clear) matrix, according to the Maxwell-Garnett effective medium approximation \cite{Maxwell-Garnett1906}. The formation process of these metallic domains may consist of  an insulator-to-metal transition similar to the one observed during the dehydrogenation  of YH$_3$, \cite{StepanovBourGartzEtAl2001,LeeKuoLinEtAl2001} according to the reversible drop of the electrical resistivity  observed in the  films during illumination \cite{MongstadPlatzer-BjoerkmanMaehlenEtAl2011}. Our results also support a recent work by Chandran \textit{et al.} \cite{ChandranSchreudersDamEtAl2014} which reported  changes of the hydrogen species in oxygen-containing yttrium hydride after illumination, suggesting the release of electrons and the formation of a metallic phase.

\begin{figure}
\centering
 \includegraphics[width=0.45\textwidth]{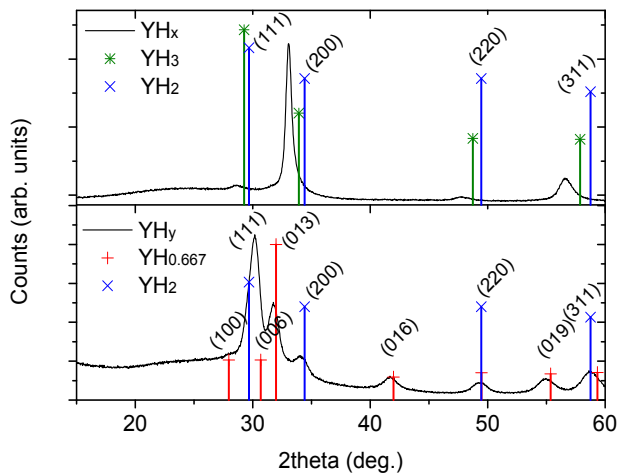}%
\caption{GIXRD patterns for transparent-semiconducting YH$_x$ [upper panel] and opaque-metallic YH$_y$ [lower panel] thin films. The standard patterns (see text for references) of \emph{fcc} YH$_2$ and high pressure-\emph{fcc} YH$_3$, as well as for hexagonal YH$_{0.667}$ have been included for comparison.}%
\label{Fig:1}
\end{figure}

Figure \ref{Fig:1} shows grazing incidence X-ray diffraction (GIXRD) patterns -- obtained by using  Cu-K$\alpha$ radiation at a fixed angle of incidence of 2$^\circ$ in a Bruker Siemens D5000 diffractometer -- for  both transparent-semiconducting YH$_x$ [upper panel] and opaque-metallic YH$_y$ [lower panel] thin films. 
The diffractogram for the  YH$_y$ thin films  correspond to a mix of \emph{fcc}-YH$_2$ and hexagonal-YH$_{0.667}$ metallic phases, according to the standard patterns JCPDS 04-006-6935 and JCPDS 01-074-8440. The main peak, observed at around $30^\circ$ is formed by the contribution of the YH$_{0.667}$ and YH$_{2}$ phases in the (111) and (066) directions, respectively. Unfortunately, the analysis of the GIXRD pattern corresponding to the transparent-semiconducting YH$_x$ films  deposited at higher $\Gamma$ is not  straightforward. These films exhibit the characteristic features of an \emph{fcc} crystalline structure that can be attributed either to an YH$_2$ or to  an YH$_3$  (JCPDS 04-015-2399) phase. None of these two possibilities provide a satisfactory explanation to the GIXRD pattern of the YH$_x$ films; as stated before, both \emph{fcc}-YH$_2$ and \emph{fcc}-YH$_3$ phases are metallic, while the YH$_x$ films are semiconducting and transparent. In addition, the \emph{fcc}-YH$_3$ is stable only at high pressure. However, theoretical studies have demonstrated that a wide-band gap \emph{fcc} phase, i.e. transparent-semiconducting, can be achieved at ambient conditions from the oxygenation of  YH$_3$ or YH$_2$ films \cite{PishtshevKarazhanov2014}. Since yttrium hydride films deposited by sputtering and exposed to air are known to oxidize heavily (see discussion below), the diffractogram measured for the YH$_x$ films  cannot be  unambiguously attributed  to an YH$_3$ or YH$_2$ stoichiometry. A complete analysis of the X-ray diffractograms for these samples can be found in the early works carried out in our laboratory  \cite{MongstadPlatzer-BjoerkmanKarazhanovEtAl2011,MaehlenMongstadYouEtAl2013}. According to  previous experiments, a lattice contraction of $\sim$ 0.3 - 0.4 \% was observed in these YH$_x$ films upon illumination \cite{MaehlenMongstadYouEtAl2013}.

Despite this uncertainty on the  stoichiometry of the  YH$_x$ films, the GIXRD analysis provide two important conclusions. On the one hand, as expected, the diffractograms confirm that the samples deposited at lower $\Gamma$ contain less hydrogen, i.e.  $x>y$, because they exhibit an hydrogen-deficient hexagonal phase (YH$_{0.667}$).   
\begin{figure}
\centering
\includegraphics[width=0.48\textwidth]{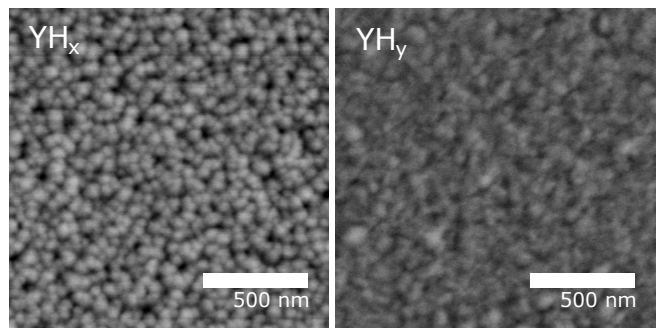}
\caption{ SEM micrograph for transparent-semiconducting YH$_x$ [left panel] and opaque-metallic YH$_y$ [right panel] thin films.}%
\label{Fig:2}
\end{figure}
On the other hand, the  lattice parameter corresponding to the \emph{fcc} cubic phase matches very well the YH$_2$ standard of 5.20 \AA~in the case of the YH$_y$ films, but it is  much higher, about 5.40 \AA, for the YH$_x$ samples. As a result, the experimental diffraction peaks for YH$_x$ in the directions (311), (220), (200) and (111) are displaced towards smaller angles compared to the standard, see Figure\ref{Fig:1} [upper pannel]. The same peaks, however, are located at the expected positions according to the standard in the YH$_y$ films, Figure \ref{Fig:1} [lower panel]. The increase of the lattice parameter in the YH$_x$ films has been observed before  and attributed to a high oxygen content in the lattice \cite{MongstadPlatzer-BjoerkmanKarazhanovEtAl2011}. Therefore and according to these observations, the oxygen content in the YH$_x$ is expected to be much higher than the one in the YH$_y$. Indeed, an energy-dispersive X-ray spectroscopy analysis (EDS) of the equivalent samples deposited onto carbon substrates, revealed an oxygen-yttrium atomic ratio of 1.29 and 0.40 for the YH$_x$ and YH$_y$ respectively. The reason why the films deposited at lower $\Gamma$ contain less oxygen still unclear, but it can be tentatively explained by using the micrographs  obtained by scanning electron microscopy (SEM) shown in Figure \ref{Fig:2}. The samples deposited at higher $\Gamma$ present a porous structure -- Figure \ref{Fig:2} [left panel] -- while the films deposited at lower $\Gamma$ exhibit much more compact features -- Figure \ref{Fig:2} [right panel] --. After the deposition process, when the samples are exposed to air, the oxidation is known to take place through the pores and holes of the film \cite{StepanovBourGartzEtAl2001}. Both the micrographs and the EDS analysis were carried out in a Hitachi S-400 scanning electron microscope.

\begin{figure}
\centering
\includegraphics[width=0.45\textwidth]{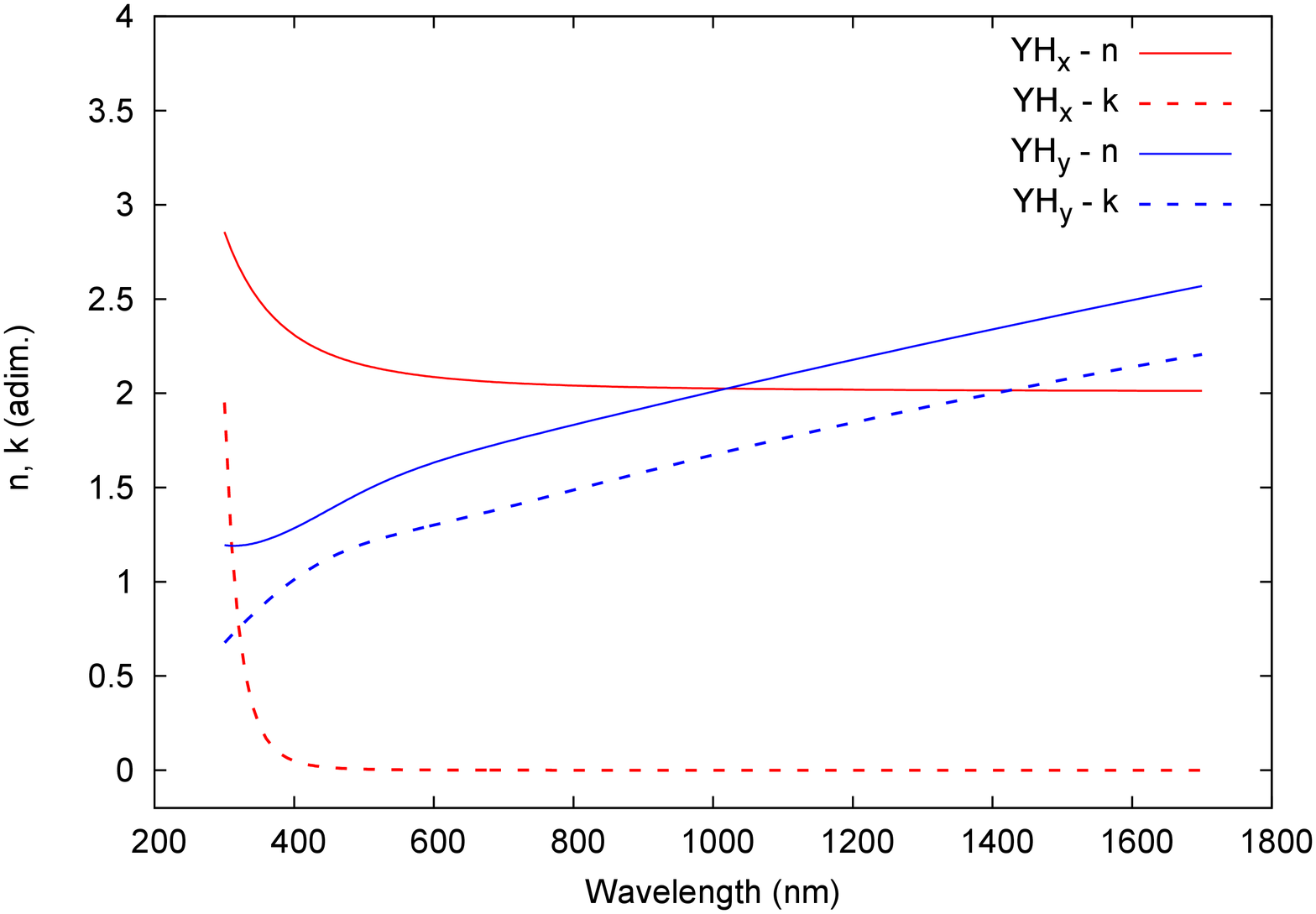}
\caption{Real component $n$ (solid lines) and imaginary component $k$ (dashed lines) of the complex refractive index, $\tilde{n},$ for semiconducting-transparent YH$_x$ (red color) and metallic-opaque YH$_y$ (blue color) thin films.}%
\label{Fig:3}
\end{figure}

Figure \ref{Fig:3} shows the real component, $n,$ and imaginary component, $k,$ of  the complex refractive index $\tilde{n} = n + ik$ for the YH$_x$ and  YH$_y$ films as obtained by variable angle spectroscopic ellipsometry. The determination of $\tilde{n}$ for each film was achieved, as usual, by the fitting of the experimental ellipsometric angles $\Delta$ and $\Psi$ to a theoretical model (not shown).  In this work, both modeling and data fitting were performed by using the commercial software WVASE32 from J.A. Woollam Co., Inc. This software performs an iterative fitting to the experimental data by using the Marquardt-Levenberg algorithm \cite{Kelley1999,Tompkins2006}. In the case of the YH$_x$ films, a simple Cauchy model \cite{Klingshirn2005} combined with an Urbach tail (exponential) provided a good fitting to the experimental measurements, while for the YH$_y$ films a Lorentz and a Tauc-Lorentz oscillators were considered \cite{Klingshirn2005}.
%
As expected, the YH$_x$ films exhibit the typical features of a wide-band gap semiconductor sample (Figure \ref{Fig:3}), i.e., no absorption ($k = 0$) in a wide range of wavelengths, including the visible; however,  below 400 nm $k$ increases exponentially, signifying the across-the-gap optical absorption. On the other hand, $n$ varies slowly as a function of the wavelength, an usual behavior of a wide range of  dielectrics and semiconductors \cite{Palik2012}. Further details on the band gap variation in these films can be found in the literature \cite{YouMongstadMaehlenEtAl2014}.
As expected, the metallic YH$_y$ films exhibit high absorption in the whole region of the spectrum studied (Figure \ref{Fig:3}).

The ellipsometric measurements revealed a film thickness of 690 nm for the YH$_x$ films, which agreed very well to the result obtained by profilometry (698 nm). On the other hand, the big optical absorption in the YH$_y$ made difficult the determination of the film thickness by ellipsometry; the thickness of these samples is of about  580 nm as measured by profilometry.

\begin{figure}
\centering
\includegraphics[width=0.46\textwidth]{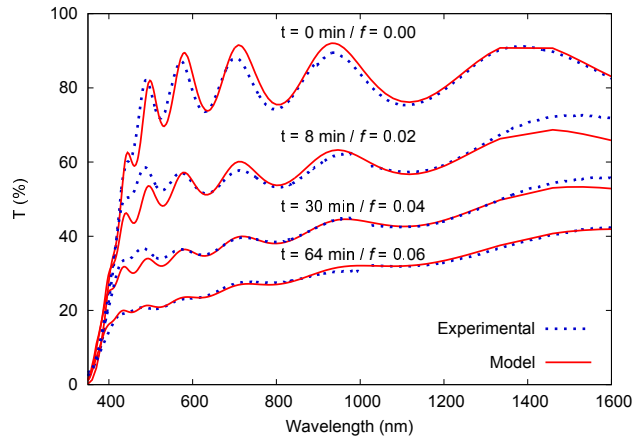}
\caption{Transmittance of the YH$_x$ films after  0, 8, 30 and 64 minutes of illumination (blue dashed lines). The transmittance calculated for  composite films with filling factors $f =$ 0, 0.02, 0.04 and 0.06 have been also included for comparison (solid red lines). }%
\label{Fig:4}
\end{figure}

Figure \ref{Fig:4} shows the evolution of the experimental optical transmittance (T) vs. time (t) of the YH$_x$ films under illumination (solid red lines). The samples were illuminated by a broad band light source EQ-99XFC LDLS with intense UV component, and the transmittance in the 300-1700 nm range was measured \emph{in-situ}  by an Ocean Optics QE65000 and  a NIRQUEST512 spectrophotometers equipped with a integrating sphere. 
In addition, the calculated transmittance of the  films -- in the clear state at different stages of photodarkening -- has been included in Figure \ref{Fig:4} (dashed blue lines). The transmittance curves has been calculated by using the Fresnel equations for a thin film over a non absorbing substrate \cite{BornWolfBhatiaEtAl2000,BhattacharyyaGayenPaulEtAl2009} assuming: \emph{(i)} the YH$_x$ (photodarkened) films consist of a composite of YH$_x$ (clear) and YH$_y$ phases, and \emph{(ii)} this composite behaves optically like an homogeneous medium of \emph{effective} dielectric permittivity $\tilde{\varepsilon}_{e\! f\! f}$.  Assumption \emph{(ii)} is valid only if the size  of the YH$_y$ inhomogeneity domains  are much smaller than the wavelength of the incident light \cite{Aspnes1982,NiklassonGranqvistHunderi1981}.
On this basis, the optical constants of the photodarkened films can be approached by using Maxwell-Garnett effective medium theory, which is suitable for  composites constituted by  nanoparticles dispersed in a continuous matrix in the dilute limit \cite{NiklassonGranqvistHunderi1981}.
According to the Maxwell-Garnett theory, $\tilde{\varepsilon}_{e\! f\! f}$ is given by \cite{Maxwell-Garnett1906}:
\begin{equation}
\tilde{\varepsilon}_{e\! f\! f} = \tilde{\varepsilon}_{{\mathrm{YH}}_x}\frac{1+\frac{2}{3}f\alpha}{1-\frac{1}{3}f\alpha}
\label{eq:2}
\end{equation}

\noindent being $\alpha$ a parameter such as:

\begin{equation}
\alpha = \frac{\tilde{\varepsilon}_{{\mathrm{YH}}_y}-\tilde{\varepsilon}_{{\mathrm{YH}}_x}}{\tilde{\varepsilon}_{{\mathrm{YH}}_x}+L(\tilde{\varepsilon}_{{\mathrm{YH}}_y}-\tilde{\varepsilon}_{{\mathrm{YH}}_x})}
\label{eq:3}
\end{equation}

\noindent   where $\tilde{\varepsilon}_{{\mathrm{YH}}_x}$ and $\tilde{\varepsilon}_{{\mathrm{YH}}_y}$  are the respective dielectric functions of the YH$_x$(clear) and YH$_y$ phases, $f$ is the filling factor (i.e., volume fraction) of the YH$_y$ phase and $L$ the depolarization factor. In this work  the inhomogeneity domains have been considered to be spherical, therefore $L=1/3$. On the other hand, $\tilde{\varepsilon}_{{\mathrm{YH}}_x}$ and $\tilde{\varepsilon}_{{\mathrm{YH}}_y}$  can be obtained  from their respective experimental complex refractive index presented in Figure \ref{Fig:3} \cite{BornWolfBhatiaEtAl2000}.
As expected, the transmittance calculated for $f=0$ (i.e. $\tilde{\varepsilon}_{e\! f\! f}=\tilde{\varepsilon}_{{\mathrm{YH}}_x}$) matches very well the transmittance of the YH$_x$ film in the clear state ($t = 0$ min). 
Under illumination, the transparency of the films decreases gradually with time; the experimental transmittance of the YH$_x$ film  after 8, 30 and 64 minutes of illumination corresponds to the one calculated for a YH$_x$(clear)/YH$_y$ composite film with $f =$ 0.02, 0.04 and 0.06 respectively (the dilute limit holds for these values of $f$). According to Figure ~\ref{Fig:4}, small inclusions of the metallic phase YH$_y$ in the YH$_x$ (clear) matrix may cause big changes in the total transmittance of the film. In fact, only a 2\% in volume of the metallic YH$_y$ phase ($f=$0.02) may cause a drop of the transmittance in the visible region in more than a 30\%. 
%
%

\begin{figure}
\centering
\includegraphics[width=0.45\textwidth]{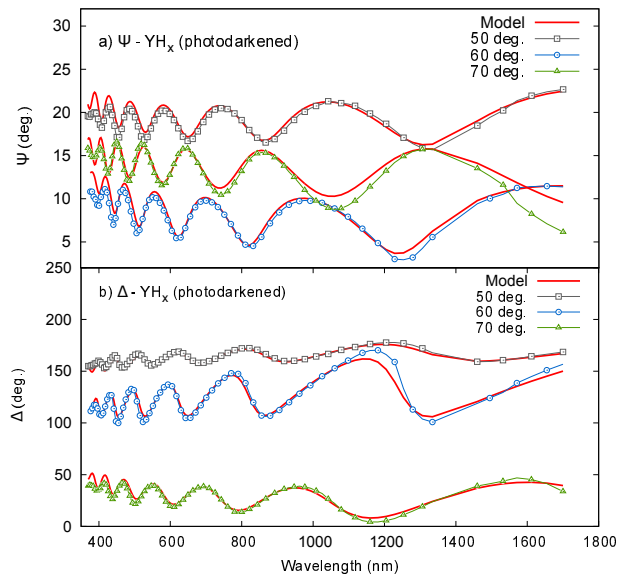}
\caption{ Ellipsometric angles $\Psi$ and $\Delta$ as a function of wavelength. These data were collected at three angles of incidence (50$^\circ$, 60$^\circ$ and 70$^\circ$) in YH$_x$ films under violet-light laser illumination. Both experimental (lines with symbols) and the model fit (solid curves) are presented.}%
\label{Fig:5}
\end{figure}

The effective medium model also reproduces very well $\Psi$ and $\Delta$. Figure \ref{Fig:5} shows  $\Psi$ [panel \emph{a}] and $\Delta$ [panel \emph{b}] corresponding to the photodarkened sample YH$_x$, measured for three different angles of incidence, 50$^\circ$, 60$^\circ$ and 70$^\circ$. The experimental results corresponds to the photodarkened film ($f = 0.06$), and are depicted by means of lines with symbols, while the best fit is represented in solid red lines. 
The calculated experimental ellipsometric spectra has been done assuming a thin film of optical constants described by  Equation \ref{eq:2} and assuming $f = 0.06$.

In summary, two sets of samples, labeled as YH$_x$ and YH$_y$,  were  deposited  at high and low hydrogen concentrations onto glass substrates.
The YH$_x$ films exhibit an \emph{fcc} structure similar to the one observed in YH$_2$ but with bigger lattice parameter $\sim 3.40$ \AA. These films are transparent, photochromic and contain a high oxygen concentration (oxygen/yttrium ratio of 1.29). 
On the other hand, the YH$_y$ films were found to be opaque, exhibit a metallic behavior and are composed by YH$_2$ and YH$_{0.667}$ phases. The oxygen content in  the YH$_y$ films is smaller that in the previous case (oxygen/yttrium ratio of 0.40).
The dielectric permittivity in the  300 and 1700 nm wavelength range was determined by ellipsometry for both sets of samples,YH$_x$ and YH$_y$. 
On the other hand, the  study of the photodarkened YH$_x$ films revealed that their optical properties  correspond to a composite of both phases, YH$_y$/and YH$_x$ (clear), calculated according to Maxwell-Garnett theory. In other words, the optical properties of the photodarkened films can be explained quantitatively by the gradual formation of small  metallic domains in the semiconducting matrix due to the action of energetic light.  
It is important to notice that the formation of a small domains  of the metallic phase (with a volume fraction of $\sim$ 0.06 or smaller) is able to cause a substantial decrease in the optical transmittance of the films. 
Although these results provide valuable hints, further investigations are needed in order to reach a complete understanding of the photochromic mechanism in oxygen-containing yttrium hydride thin films.

This work has been supported by the Norwegian Research Council through the FRINATEK project 240477/F20.


\end{document}